\documentclass[11pt, titlepage=false]{scrartcl}

\usepackage[utf8]{inputenc}

\usepackage{type1cm}        
\usepackage{makeidx}         
\usepackage{graphicx}        
\usepackage{multicol}        
\usepackage[bottom]{footmisc}

\usepackage{newtxtext}       %
\usepackage{newtxmath}       

\makeindex             

\PassOptionsToPackage{hyphens}{url}
\usepackage{hyperref}
\usepackage[anythingbreaks]{breakurl}

\begin{document}

\title{Approaching Ethical Guidelines for Data Scientists }

\author{\small{ Ursula Garzcarek} \\ \small{ Cytel Inc, Clinical Research Services
  ICC}\\ \small{Route de Pré-Bois, 20 C.P. 1839, 1215 Geneva 15, Switzerland} \\
  \small{Ursula.Garczarek@cytel.com}  
  \and \small{Detlef Steuer} \\\small{ Helmut-Schmidt-Universität, Universität der Bundeswehr Hamburg} \\ \small{Holstenhofweg 85, 22043 Hamburg, Germany}
  \\ \small{steuer@hsu-hh.de}}

\date{ \small{ Last updated \today}}

\maketitle


\begin{abstract}
The goal of this article is to inspire data scientists to
 participate in the debate on the impact that their professional work
 has on society, and to become active in public debates on the digital
 world as data science professionals. How do ethical principles (e.g., fairness, 
justice, beneficence, and non-maleficence) relate to actual situations in our
 professional lives? What lies in our responsibility as professionals
 by our expertise in the field? 
More specifically this article makes an appeal to statisticians that may consider themselves not as data scientists, 
nor what they do as data science, to join that debate, and to be part of the community 
that establishes data science as a proper profession in the sense of Airaksinen \cite{TimoAiraksinen}, a philosopher working on professional ethics. 
As we will argue, data science has one of its roots in statistics and at the same time extends beyond it. To shape the future of statistics, 
and to take responsibility for the statistical contributions to data science, statisticians should actively engage in the discussions.
\\
In Section \ref{definition} the term data science is defined, and the technical changes that have led to a strong influence of data science on society are outlined. In Section \ref{cnil} the systematic approach from \cite{CNIL2018} is introduced. Along the lines of that approach prominent
 examples are given for ethical issues arising
 from the work of data scientists. In 
Section \ref{guidance} we provide reasons why data scientists 
should engage in shaping morality around data science
 and to formulate codes of conduct and codes of practice for data
 science professionals.
 In Section \ref{existing}  we present established ethical guidelines for the related fields of statistics and computing machinery. Section \ref{devel} describes necessary steps in the community to develop professional ethics for data science. Finally in Section \ref{conclusion} we motivate our own engagement and give our starting statement for the debate:
 \emph{Data science is in the focal point of current societal development. 
Without becoming a profession with professional ethics, data science will fail in building trust in its interaction 
with and its  much needed contributions to society!}
\end{abstract}

 \section{Definition of data science, the roots and the changing role}
 \label{definition}
We start with the definition of data science as given by Donoho which we find very useful. We will describe how data science relates to statistics and machine learning and why the role of a data scientists in society is becoming increasingly important.

\subsection {Definition of data science}
\label{sec:defdatascience}
There is currently no generally agreed definition of data science. Here we use the definition of Donoho \cite{Donoho} of \emph{greater} data science:
\\
\emph{Data science is the science of learning from data; it studies
the methods involved in the analysis and processing of data and
proposes technology to improve methods in an evidence-based manner.
The scope and impact of this science will expand enormously in coming
decades as scientific data and data about science itself become ubiquitously available.}

Donoho also provides a classification of the related activities into six divisions:
\begin{enumerate}
\item Data gathering, preparation, and exploration,
\item data representation and transformation,
\item computing with data,
\item data modeling,
\item data visualization and presentation,
\item science about data science.
\end{enumerate}
Items 1 to 5 describe the work of a data scientist, 
item 6 differentiates what he calls \emph{greater data science} from data science.

\subsection{Relation of data science, statistics and artificial intelligence}
The lack of an agreed definition of data science is a symptom of a larger problem: it is not (yet) a profession of its own. 
Some see it as subdivision of machine learning, and thus a subdivision of artificial intelligence, others as subdivision of statistics, that is exploratory statistics, 
and many see it as a collection of methods from both statistics and machine learning,  used by people of different professional backgrounds, or people with no actual 
professional background only trained in the application of those methods, without the necessary formal scientific education. By starting with the definition of Donoho (sec. \ref{sec:defdatascience}) we already make two statements: 
\begin{enumerate}
\item Data science should become a profession in the sense of Airaksinen \cite{TimoAiraksinen}, with a definition, a grounding in science, and a task and responsibility in society, and
\item exploratory statistics is a historical predecessor of data science.
\end{enumerate}
With respect to the second point, we do not claim exploratory statistics to be the only predecessor of data science. With the same right, people from the artificial intelligence community can see machine learning as a historical predecessor of data science. Therefore, we want the machine learning and artificial intelligence community to work together with the statistics community on the first point.

\subsection{The changes of the societal impact of data science}

The biggest, relatively recent changes in practical data science are the availability of vast amount of data together with the increase
 in computational power. Technically speaking this enables fast, low-cost processing of ever-changing large data bases by algorithms to derive continuously updated highly condensed and aggregated data, i.e.\ results. 
These results can be fed into human decision making, that is based on the interpretation and understanding of the results, or they can be used in rules for automatic decision making. Whether or not, at least interim, the decisions are made with human understanding of the results and how they were generated, distinguishes black-box algorithms from other algorithms.

Focus of this article are the consequences of processing and analysing vast amounts of data about humans and human behaviour. Todays possibilities in these respects change human interaction and thus society directly and fundamentally. 
Examples for this broad claim will be given in subsequent sections. 

As data science is the focal point of these developments the role of data scientists in society becomes more influential and important. With increased influence and importance comes increased responsibility.

\section{Ethical issues in data science}

The awareness that data science and its algorithms have an increased and fundamental impact on society is vivid around the world. 
There are ongoing or starting discussions in many countries and organisations in legal and political context, actually too many to cite. Instead, we refer to any search in news portals, social media and internet with terms as \emph{algorithm, impact, society}.

Actually such considerations are not really new.
To our knowledge, the first data science application recognised to have a large impact on societal processes are election forecasts and polls on voting behaviour.
Many countries have thus regulations on what is allowed to publish when in context of an upcoming or ongoing elections. An overview over such regulations is given in  \cite{sperrfrist:wahlen}.

A systematic approach to identify, describe and categorise  those ethical issues was undertaken by CNIL (Commission nationale de l'informatique et des libertés) in 2017 \cite{CNIL2017, CNIL2018}.
The report is the result of a public debate organized by the french data protection authority. We will follow its structure and give examples for each of the given categories of ethical issues to make them tangible.
The main points relevant for consideration by data scientists are identified.

\subsection{ Six main ethical issues according to CNIL}
\label{cnil}
In the debate six main issues were identified. Citations referencing \cite{CNIL2017} are given in front of each of the following sections. These ciatations are set in italics to be easily identifiable.

\subsubsection{ Autonomous machines: a threat to free will and responsibility?}
\label{auto}
\emph{Delegation of complex and critical decisions and tasks to machines
increases the human capacity to act and poses a threat to human
autonomy and free will and may water down responsibilities.}

The most widely discussed application of this type are autonomous
vehicles. Autonomous vehicles have the potential to increase traffic safety, but who is responsible
for remaining accidents? Will it be possible to overrule a machine's decision on
\emph{lowest} or \emph{allowable} risk, i.e.\ in case of an emergency.

On a more abstract level 
any sufficiently complex system may be called an autonomous machine.

Already today many Kafkaesque situations arise due to complex semi-automatic regulations, i.e.\ the story of a man who was released from his job by an algorithm due to an error, and no human
was able to stop that procedure \cite{laidoff} after the lay-off was triggered.

It must be noted, that in these settings the 
data scientist is not involved directly. May be she or he built some model in preparation 
to steer the machine, but the implementation generally was not her or his task.

\subsubsection{Bias, discrimination and exclusion} 
\label{bias}
\emph{Algorithms and artificial intelligence can create biases, discrimination or even exclusion towards individuals and groups of people}

\paragraph{General remarks} 

This issue is one where data science expertise is very important for understanding the extent of the problem. 
We start stressing one point that is often overlooked, when algorithmic bias is discussed. 
The very nature of the most commonly applied algorithms, -called pattern recognition or classification and clustering-,
if applied to humans, is applying \emph{prejudice}. In statistical language they form a prior belief on an individual generated by
experience with other individuals assigned to the same group. Goal of these algorithms is the assignment of a new object, in this case a person, 
according to some measured characteristics of this person into some group. Judgements and predictions on e.g. future 
behaviour or reactions to a medical treatment for the individual are then made according to previously observed behaviours 
or reactions of the other's in the group. Obviously, if this leads to an improved medical decision making, this is to the 
benefit to the individual and the society at large. 

In many examples, though, there is a possible benefit to some and a negative impact on others. In those cases, questions of fairness and justice are touched by the use of these algorithms for judgement/prediction and decision making in general.
Any of their use constitute \emph{bias}, if the measured characteristics, that lead to the assignment into the group, are only 
correlated but not causally related to the features that are judged about. Formally the reason is, that the relationship 
between what is predicted or judged about for the individual and the measured characteristic of the individual is 
conditionally independent given the individual. Note that this bias is created independently on whether or not the 
underlying database is representative for the larger population for the measured characteristics. The bias is created by 
applying an approach (= data + method) that is suitable for correlational analyses only for judgements that require causal reasoning on individual level. 

Practically this is not different from humans basing their judgement on a person, on experiences (= data) they have made with 
other people that are alike based on some arbitrary (that is bearing no causal relationship) assessment on similarity. 
If this is implemented by an algorithm the impact can be more severe, as the identical bias is applied to more people and 
forms a more systematic bias towards certain groups.
Combined with monopolies on data ownership, - like currently for social media or search data -, and 
with the scalability of computing power such a systematic bias can easily become a universal norm.
Where the algorithm uses characteristics that  include or are related to protected characteristics by anti-discrimination 
laws (mostly race, sexual orientation, religion or belief, age and disability) any judgement and any decision based on the 
algorithm constitute instances of \emph{discrimination}, when they result in one person being treated less favourably than 
another in a comparable situation.

This does not happen only in badly designed or malfunctioning systems. It is in the core of all classification applied to 
people.

Another, -practically incurable-, drawback of those algorithms is that they infer from 
data of the past, - on the members of the group and/or the individual on which one wants to judge -, and human behaviour on an individual level and
their patterns do change over time.

\paragraph{Examples} 

The probably most famous example is COMPAS (Correctional Offender Management Profiling for Alternative Sanctions) a software used in the US judicial system to
classify the probability of defendants' recidivism. 
A good discussion of the approach can be found in \cite{ONeil}. It was shown in a detailed analysis  \cite{propublica, propublica2} that
the privately owned algorithm used in the juridical system gave far better prognoses for
white than for black people, thus it discriminated implicitly based on color. The machine generated prognosis was intended just to help the judges,
but in interviews it could be seen, that it played a crucial rule in the judgements. Especially decisions by the judges 
whether defendants could get out on parole or had to go to jail were strongly influenced by the algorithm's output and discriminated against black people.
 
It must be stressed, that this bias in application was not intentional as far as it is known. The bias most probably was introduced through available data on prisoners in conjunction with the above described fundamental misunderstanding that observed correlations would be good enough to make decisions that require causal reasoning.

Examples of the application of algorithms are not restricted to the US. In Europe for example there is a recent
initiative in Austria to classify unemployed people in one of the three possible groups:  
\emph{bad (<= 25\%)}, \emph{mediocre} or \emph{good chances (>=66\%)} to be employed for at least 6 months in 24 months from now \cite{oesterreich}.
The idea is to spend money to bring people back into the workforce more on target.
Controversial is the stated goal to spend \emph{less} money on those in the lowest group.
It is reported that age and nationality increase one's probability to be put in the lowest group.
Both points seem to be openly discriminating. The official stance is, that the algorithm does not decide,
but only helps a human to decide and therefore no discrimination would happen. This is ignoring to the large influence that those 
supportive systems have, when there is a shortage of money: decision makers typically need to justify, if they deviate from the 
algorithmic choices, but not if they follow the machine's decision.  The default mode of operation may change through the use of such a simple helper
algorithm.
\\
A very similar system is already in use in Poland \cite{panoptykon}.

In the examples given, in addition to generating bias, the automatic classifiers act like self-fulfilling prophecies.
The automatic, even secret, classification of an individual will influence his or her future life, in the direction 
the chosen algorithm determines. 
At the same time it becomes impossible to assess the algorithms performance in the future, as the future of the individual's life is changed based on the algorithms outcome and there is no control group.

Also the algorithms act very similar to ancient oracles. For an outsider it is impossible to find out
which characteristics of a person exactly have led to the given classification. They are black-box algorithms, a feature shared by many of the algorithms from the artificial intelligence community. There only is the saying of the oracle, no reasoning, and no possible
recourse. Black-box algorithms therefore will always be problematic for usage in any juridical system or for any scoring implying a value judgement of an individual, i.e.~ credit scoring.

These applications are examples for applications where some people have a benefit and others negative
 consequences from the application of the algorithm. It is accepted, that the application may be not in the
 interest of the individual that is judged.

Of course this is not a drawback inherent in using algorithmic decision making.
It is possible to set up procedures with no intention to inflict negative consequences on some to the benefit of others, if care is given to transparency and possible discriminating behaviours. 
For example in Germany there exists a program RADAR-iTE (Regelbasierte Analyse potentiell destruktiver Täter zur Einschätzung des akuten Risikos - islamistischer Terrorismus)  \cite{RADAR-iTE} where an algorithm is used to try identifying the more
dangerous people in a group of people already under investigation by law enforcement. 

Decisions are based on a set of 72 questions which are transparent for anybody involved. Because those under
inspection by RADAR-iTE already are under investigation, the most important aspect of its application is resource allocation
by law enforcement. There is no additional negative effect on those individuals that are judged to be high
 risk beyond being under investigation already. Publicized numbers \cite{faz:radar} give around half 
(96 of 205) of the suspects are considered \emph{low risk} after
classification by RADAR-iTE, only around 40\% (82 of 205) are considered \emph{high risk}.
Transparency of all steps seems guaranteed throughout all decisions performed with respect to algorithmic classifications.

In this case those applying the algorithms and those being judged share in some sense
the goal to reduce the number of individuals that are observed. The application of the algorithm
has the potential to help an individual by being removed from the group of \emph{high risk} people.

The implications of a similar algorithm if it was applied to screen the overall population would lead to a completely different asessment.
Technically, there is no barrier to such a use. It can only be prevented by \emph{morality} and \emph{law}.

\subsubsection{Algorithmic profiling} 
\label{prof}
\emph{Personalizing versus collective benefits: Individuals
have gained a great deal from profiling and ever finer
segmentation. This mindset of personalising can affect the key
collective principles like democratic and cultural pluralism and
risk-sharing in the realm of insurance.} 

The most discussed form of personalizing in the age of the internet is the so-called
\emph{filter bubble} \cite{Pariser:FilterBubble}. The scandal around Cambridge Analytica using
Facebook data for micro-targeting a very specific subset of the public with the aim to
influence the US elections in 2016 made the dangers of highly personal news and marketing feeds
obvious  \cite{CambridgeAnalytica, CambridgeAnalytica2}. 

As a reaction the legislative started to formulate laws to reduce the risks of such personalized
targeting with fabricated news, i.e. in Germany the ``Netzwerkdurchsetzungsgesetz'' \cite{NetzDG}.
Facebook restricted the admission to personal data for
third parties in the aftermath of that scandal \cite{facebook:restrictions}.

A data scientists role, if
implementing schemes for targeting specific sub-population identified by profiling with the
help of the vast amount of information available on each active person in the internet,
should at least be to warn of possible  misuse.
She or he should understand the \emph{dangers for society} and only help to implement \emph{lawful or ethical}
algorithms.

A nice example for the second point on risk-sharing are telemetry data collected by so-called smart devices and transmitted
to insurance companies. Since the beginning of 2018 each new automobile in the EU has to record telemetry data
in a system called eCall \cite{eCall}. While that system will only transfer data in case of an emergency, there are systems that collect lots of information
about all aspects of car usage, down to location and the music the driver listens to \cite{cardata}. 
First there are obvious problems with privacy, if there can be unlawful information sharing. The second problem
here are insurance companies who try to give personalized policy premiums based on level of data sharing a car owner
accepts. Probably even more problematic are health data, which can be accessed by insurance companies \cite{forbes}.

While at first nothing seems at stake if an unhealthy living style is punished with higher policy costs, a second look
reveals that the fundamental principle of an insurance, namely risk sharing among a large group, is eroded. In addition 
there is a direct conflict of personalized insurance policies and personal freedom. 
Big monetary pressure on customers to live a good live \emph{in the sense of the insurance companies} must be expected.

\subsubsection{Preventing massive files while enhancing AI: seeking a new balance} 
\label{priv}
\emph{Artificial intelligence by being based on advanced techniques
of machine learning requires a significant amount of data. Still, data protection laws are rooted in the belief that individuals’ rights
regarding their personal data must be protected and thus prevent the creation of massive files. AI brings up many hopes: to what
extent the balance chosen by the lawmaker and applied until now should be renegotiated?}

A field of research that is already very experienced and advanced in using large databases on humans and trying to find ways to make 
that balance is the medical field.
Thus the following two examples are able to illustrate the benefits of the availability
of collected personal data and how 
the risks for individuals regarding their privacy or for the society regarding fair access to information were mitigated.

In July 2018 some valsartan products were discovered to have been contaminated with N-nitrosodimethylamine (NDMA). 
In September 2018 an expedited assessment of cancer risk associated with exposure to NDMA through contaminated 
valsartan products could be published \cite{Pottegard2018}, providing reassuring interim evidence that the short
 term overall risk of cancer in users of valsartan contaminated with NDMA was not markedly increased. This fast 
assessment in a relatively large cohort (5150 Danish patients) was possible by linking data from four official
 Danish registries on individual level thus collecting information on prescriptions, cancer diagnosis hospital admissions,
 mortality and migration. Privacy was implemented by a process where officials from the registries perform the linking,
 derive the important information, and then de-identify the data before it is sent to the scientists.
 
In 2018 the German health insurance company DAK Gesundheit in cooperation with scientists from the University of 
Bielefeld published a report on the health status and the health costs of children and adolescents based on the claims database 
from the people insured with the DAK Gesundheit \cite{DAK1}. Next to some general overview on the health status, a key topic was
 the investigation of the influence of socioeconomic status and education of the parents on the health and induced health
 costs of the children. The main conclusion is that education is a stronger influencing factor than socioeconomic status 
and that important preventive measures consist of giving children good health education. In the same report, and by guest
authors \cite{DAK2}, also the results from the KiGGS study \cite{KIGGS} are discussed. That study puts its emphasis more on the
principle of equal opportunity and the influence of socioeconomic status on general health and specifically mental health. 
Publishing this together shows sensitivity of the topic in the political debate and the role that an open scientific environment has to play.

Both, the valsartan case and the DAK study show that there are true benefits for public health that can be generated from 
using large medical databases. When balancing these benefits with the risk for privacy violations for the people whose data 
is used, in the valsartan case, we want to highlight the high trust from the citizens that is given to officials: if data on 
any medical problem one encounters in life can be linked to the home address, citizens need to trust the government that this 
data is not accessible or made accessible to anyone that uses this information with other than the best intentions. With the 
DAK study we want to highlight another important aspect of balancing benefit-risk: the ownership of data, and fair access to 
data. Data is the \emph{new oil}, and evidence generation shapes how benefit is defined and how it is implemented. Thus, if risk is 
shared by people of all political opinions, then fairness requires that evidence generation is possible for people from 
different political opinions.

In general, an important measure for respecting privacy is to de-identify data in the databases, and making them 
non-identifiable. Guidelines exist for de-identification processes (e.g. the Safe Harbor method \cite{HIPAA}), yet, with 
growing databases through social media use and genetic and biomarker research, non-identifiability is a moving target. 
A good counter-measure is implemented in the process for requesting access in the so called MIMIC-III database 
\cite{Goldbergeretal} on critical care unit patients. In addition to a required training on data privacy, and a strict 
de-identification of the data, all scientists accessing the data have to submit a data use agreement with 10 points, 
among which there is one requiring the scientists take immediate action should they realize that there is a way to 
de-identify data. This is acknowledging the fact that de-identification is no guarantee to de-identifiability at all 
times by installing a process to monitor de-identifiability by those who have the expertise and knowledge, namely the data scientists, holding 
them responsible for it and giving them, as a community, a general credit of trust.

\subsubsection{ Quality, quantity, relevance: the challenges of data curated for AI}
\label{qual}
\emph{The acceptance of the existence of potential bias in datasets curated to train algorithms is of
paramount importance. }

Even if implemented in best of mind, there may be unexpected bias in the training data going beyond what has already 
been said about bias in Section \ref{bias}. There are many examples to find, we want to give two.

One famous example of algorithmic training going wrong was Microsoft's twitter bot \emph{Tay} \cite{MS:Tay}.
Tay was implemented to act on Twitter as a regular user. The bot should learn from the comments by others how to perform common twitter conversations. In less than a day the humans had learned
how to manipulate the learning algorithm in such a way that \emph{Tay} started to speak out fascistic and racist
paroles. Microsoft decided to take \emph{Tay} offline less than a day after it started learning.

A recent example for a similar event is an AI system at amazon. That system should help to find the most
qualified applicants in their huge stream of applications. The experiment had to be stopped,
when it was noted that the algorithm systematically downgraded applications of women. In \cite{reuters:amazon}
some probable causes for that behaviour are given. The training data contained mostly applications of men, so
most of the successful applicants were men. There are not too many details, but as a consequence any
appearance of the word \emph{woman} reduced the chances of that applications. 

Finally the whole project was stopped, even after the developing team tried to correct for known shortcomings, 
because \emph{there was no guarantee the machine would not devise ways to discriminate in other ways} \cite{reuters:amazon}. 

The important observation in both cases is, that these black-box algorithms couldn't be improved.
They had to be taken offline and completely replaced. As an obvious consequence such algorithms
should not be used, where such a replacement is complicated or dangerous.

\subsubsection{  Human identity before the challenge of artificial intelligence}
\label{robo}
\emph{Hybridisation between humans and machines
challenges the notion of our human uniqueness. How should we view the
new class of objects, humanoid robots, which are likely to arouse
emotional responses and attachment in humans?} 

This point from the debate in France run by CNIL is given only for the sake of completeness. At the moment, 
we do not believe that this is an ethical issue where data scientists have a special responsibility due to their expertise.

\subsection{Conclusion from CNIL's report}

The given examples show the multitude of complex ethical issues that arise from a data scientist's  work.
In the next section we argue that ethical guidelines for data scientists are one mean to help them
taking their responsibility.

\section {Guidance for data science}
\label{guidance}
The call for more guidance for digital technologies in media in general is loud and all across the globe, leading to 
various initiatives and groups engaging in discussions around ethical rules for developing and implementing those 
technologies. For an overview on initiatives and ethical values in the tech field visit the website of the think tank 
doteveryone \cite{James2018} or  the blog of Erickson \cite{Erickson2018}. There is a long history of computer scientists 
discussing the ethics of algorithms. A good starting point is the website \url{fatml.org}. Here fatml is an acronym for 
\emph{Fairness, Accountability, and Transparency in Machine Learning} and stands for a series of conferences. For the german speaking communities, we recommend the slides to the one day workshop 
\emph{Ethische Leitlinien wissenschaftlicher Fachgesellschaften} of the Deutschen Gesellschaft für Medizinische Informatik, 
Biometrie und Epidemiologie (GMDS)  \cite{gmds2017} or the Algorithmic Accountability Lab (AAL) at the University of Kaiserslautern 
\url{aalab.informatik.uni-kl.de}. AAL provides a good source for current discussions not specific for data scientists but about the use of algorithms in general with some hints toward data science.

This article is in that sense, one contribution among many. Its main purpose is to broaden the audience and increase the 
number of participants in the discussions, and to foster the development of \emph{morality}, a \emph{set of deeply held, widely shared, 
and relatively stable values} \cite{Horner} on data science within and around the data science community. As any ethical 
guidance, be it in form of codes, oaths, and even law, only has the intended impact, if people are willing to follow it, and 
the chance for that is high, if the underlying norms and values are in accordance with, in this case, the data science community's own morality. 

\subsection{Do we really need more ethical guidelines?}
Not everyone would agree that data scientists need more guidance how to make moral decisions in their professional life: many do work in companies with codes of conduct, work for institutions that require some oath, or are members of scientific societies that give ethical guidelines to their members, or have religious beliefs that give guidance to wrong or right in their life, and there is the fundamentally skeptical view that paper does not blush. Also, we are all obliged to obey to law. So what does a special set of ethical rules for the profession of data science add? 

Four rationales:
\begin{enumerate}
\item  For the individual data scientist, the translation from very general ethical principles from common morality, law or religion, to an ethical issue at work can be quite difficult. Especially since most issues are not about intentions, but about the consequences of one’s work. Those consequences are often not very easy to judge upon. Having some reference to well-thought through and well-reasoned guidelines in that sense is not more nor less than having publications on specific methods: it helps to avoid re-inventing the wheel ever so often. In addition, it can be very helpful to have such a reference along with the reasoning for justification, if the consequences of an ethical decision increase the workload for a colleague or costs for an employer or client. 

\item  For data scientists as a community, having formulated codes of conduct or some service ideal makes the 
difference of acting as professionals or merely having a job that does data crunching.  In sociology, a profession is 
defined by means of professionalism. This implies that a profession has a certain degree of autonomy in society, 
its members’ expertise is based on science, and the professional work exemplifies a service ideal \cite{TimoAiraksinen}. 
In other words: without a service ideal, there is no professionalism and without professionalism, there is no profession. 

\item  For data scientists as members of society, for their clients, employers and colleagues, written rules of conduct for data science services can help to establish a relationship of trust. If they are written clearly, they give lay people some mean to know what to expect from a data scientist, to compare what they are getting against that standard, and finally gain trust if the expectations are met. Being trusted as a professional increases social status, reputation and possibly the money that is paid for the service.

A code of conduct or ethical guidelines may even be the start of a well defined job definition for data scientist!   

\item In case of conflicts of interests an ethical guideline under the maintainership of some professional society
 may offer an arbitration process between different interests. 
\end{enumerate}
  
\section{Existing guidelines and codes}
\label{existing}
In the previous section, we provided references to ongoing efforts to develop ethical guidelines to data science itself and connected scientific or technical fields. Here, we want to give more details on the three main guidelines from the fields of statistics and computer sciences from some of the largest and oldest established associations for those communities. If one could establish additional sub-guidelines that filled the gaps with respect to data science aspects, the audience would immediately be very large, and there would be no need to establish a new association. Both, ACM and ASA, acknowledge data science as an important field in their domains.

\subsection{ASA: Ethical guideline for statistical practice}
The American Statistical Association was founded in Boston in 1839 and has more than 19000 members worldwide. 
The current Ethical Guidelines \cite{ASA:Guidelines} have been updated and approved by the ASA Board in April 2018. 
The guideline has eight sections, six of which describe the responsibilities towards individuals and groups of people 
to which the statistical work may matter:
\begin{itemize}
\item Professional integrity and accountability,
\item integrity of data and methods,
\item responsibilities to science/public/funder/client,
\item responsibilities to research subjects,
\item responsibilities to research team colleagues,
\item responsibilities to other statisticians or statistics practitioners,
\item responsibilities regarding allegations of misconduct,
\item responsibilities of employers, including organizations, individuals, attorneys, or other clients employing statistical practitioners.
\end{itemize}
Checking which of the ethical issues discussed in Section \ref{cnil} are covered, one recognises, that implicitly, it is a clear call for human responsibility addressing the issue raised on autonomous machines (Section \ref{auto}). It only touches very briefly on the risk, that information presented as aggregates on groups may lead to bias, discrimination and exclusion (Section \ref{bias}). It sets high standards for privacy and respecting data confidentiality (Section \ref{priv}). With the integrity of data and methods section and throughout almost any other point, it gives clear guidance on quality, quantity, and relevance of data, and to a general notion of scientific honesty. It also addresses ethical issues specific to human studies, not covered in section \ref{cnil}, but very relevant to all scientists working in that field. The guidelines have gaps concerning those ethical issues that result from the implementation of \emph{statistical procedures} into daily practice. Missing are discussions on all ethical issues that can arise from implementing algorithmic results without further human interaction into automatic decision making.

\subsection{ACM: Code of ethics and professional conduct}
The Association of Computing Machinery (ACM) was founded in 1947 and has more than 100.000 members worldwide.
The ACM has ethical guidelines for a long time.
\emph{The Code} \cite{ACM:Guidelines} as it is named, has just been updated and adopted by ACM in June 2018. 
It has a preamble, and four sections:
\begin{enumerate}
\item  General ethical principles,
\item professional responsibilities,
\item professional leadership responsibilities and
\item compliance with the code.
\end{enumerate}

On a general level \emph{The Code} addresses all ethical issues that we present in Section \ref{cnil}. 
Yet, the Code is not a code for data science, and it is not providing the constructive guidance ASA gives on the integrity 
of data and methods related to scientific honesty and on responsibilities to research subjects. 

\subsection{Ethical guidelines of the German Informatics Society}
The German Informatics Society (GI) has a long history of its ethical guidelines \cite{GI:Guidelines}. The latest update was in June 2018. These guidelines are concise and consist of a preamble and 12 very short sections.
\begin{itemize}
\item Sections 1 to 4 concentrate on aspects of the professional competence of computer scientists,
\item sections 5 and 6 are about individual working conditions,
\item sections 7 and 8 are about teaching and researching in the field of computer science.
\item Very interesting are sections 9, 10, and 11 which clearly state the societal responsibilities of
  computer scientists. We see some intercept with the work of data scientists there.
\item Finally section 13 defines a mediating role of the German Informatics Society in case of conflicts
  stemming from these guidelines.
\end{itemize}  
There are no data science specific sections in these guidelines, nevertheless many important aspects
are touched. We think the structure of the ethical guidelines of the GI can be a good skeleton to develop ethical guidelines for data science.

\subsection{Conclusion from examining existing guidelines}

The ethical guidelines for statisticians from the ASA are constructive and detailed for the ethical issues of 
statisticians and data scientists in the sense of Donoho (Section \ref{definition}) that work in research 
and the special responsibilities 
towards participants in human studies. \emph{The Code} of the ACM covers the area of using data from and about humans outside from human studies and issues that arise from implementing algorithms from data science for repeated use and that have impact on individuals and communities. What we have in mind is a combination of those aspects, maybe structured as in the guidelines of the GI, as data scientists work on data from all sources and across all those areas.

\section{Development of ethical guidelines for data science}
\label{devel}
There are hurdles to overcome before a meaningful guideline can be established. In our view the main ones are the lack of a sense of community and a lack of communication on ethics. 

\subsection{Data scientists have to perceive themselves as a community}

At the moment the term \emph{data scientist} in not a protected professional title. 
Data scientists can have an academic training in statistics, or computer science, as their main fields of professional training, 
but also engineering, psychology, business management, or they can be trained programmers or only have been following a three-month 
course on data science learning Python, Julia, or R. In that sense, data science today is not a profession but only an occupation. 
\cite{TimoAiraksinen}. Between the data scientists from statistics and computer science, on the ground, there is not much 
tension, but there are many turf battles on academic levels. So the first step would be to realize that ethical 
guidelines are a shared interest and to then start discussing the content within data science related societies, 
at conferences, in University courses, at work with colleagues.

Being a community does not mean that there is a need for a new association. A good option would be to add data science specific guidelines to those of the ACM, the ASA, and the GI. Such an approach would have the big advantage, that it would not require to first establish a new data science association.
Of course the authors would like to see the european statistics societies embracing ethical issues in their agenda.

\subsection{Data scientists have to overcome shyness or ignorance to discuss ethics and own moral views related to data science}

In the perception of the authors it is very uncommon for data scientists to express any moral view on the work they do or on the impact their work  may have for fellow people and the society at large. That might be, because only recently society and data scientists themselves have realized how much impact data science services have on individuals and communities. Maybe that is because  the very nature of this impact is, to be de-personalized and it is easy to overlook one's own responsibility. Maybe it is because most people in data sciences are coming from a mathematical, technical, or computer science background and are in general less vocal on anything outside hard science. 
The places to change such culture fundamentally should be universities and colleges where data science is taught. Ethics and professional ethics should be part of the curriculum, just as inspiring critical thinking and expressing one’s views. In the meantime every data scientist can work towards that goal within her or his environment. Crucial is taking part in discussions at work in critical projects or within any community when there are e.g. discussions on the so-called digital revolution, the influence of social media, or algorithms in health care or the criminal justice system.

Talking about ethical questions must become natural for any data scientist.

\section{Conclusion}
\label{conclusion}
We wrote this article for most parts without assuming that our views are
generally shared views, or that anyone has to agree that any given specific
application is good or bad. Underlying, there is an understanding that the
morality of the data science community is evolving and that it is a shared
task to develop it, which in turn needs open discussions.
Yet, there is at least one fundamental basic moral conviction of the authors, which we have taken as a generally agreed moral principle: as a human being one has to think about possible consequences of one’s actions.  That responsibility for the consequences grows with the knowledge and the potential one has to think about consequences.

Finally we want to start the the debate with a first statement:

Data science is in the focal point of current societal development. 
To build trust in data science and its interaction with society  and to empower data science to take its resposibility for its contributions to society, data science must develop professional ethics and become a clearly defined profession!


\begin{thebibliography}{99.}%
\bibitem{Donoho} David Donoho (2017), 50 Years of Data Science, Journal of Computational and Graphical Statistics, 26:4, 745-766, doi: 10.1080/10618600.2017.1384734 
\bibitem{RADAR-iTE} Bundeskriminalamt, Presseinformation: Neues Instrument zur Risikobewertung von potentiellen Gewaltstraftätern, RADAR-iTE (Regelbasierte Analyse potentiell destruktiver Täter zur Einschätzung des akuten Risikos - islamistischer Terrorismus), 2 Feb 2017, \url{https://www.bka.de/DE/Presse/Listenseite_Pressemitteilungen/2017/Presse2017/170202_Radar.html} Cited 6 Nov 2018
%
\bibitem{faz:radar} FAZ, Jeder zweite Gefährder hat das Potential zum Terroristen, 18.12.2017,\\ \url{http://www.faz.net/aktuell/politik/inland/terror-gefahr-jeder-zweite -radikale-islamist-hochgefaehrlich-15347580.html}. Cited 6 Nov 2018.
%
\bibitem{cardata} Jürgen Seeger, ADAC-Untersuchung: Autohersteller sammeln Daten in großem Stil, 4 June 2016, \url{https://www.heise.de/newsticker/meldung/ADAC-Untersuchung-Autohersteller-sammeln-Daten-in-grossem-Stil-3227102.html}. Cited 6 Nov 2018.
  %
\bibitem{propublica} Julia Angwin, Jeff Larson, Surya Mattu and Lauren Kirchner, ProPublica:  
Machine Bias, May 23, 2016, \url{https://www.propublica.org/article/machine-bias-risk-assessments-in-criminal-sentencing}. Cited 24 Oct 2018
%
\bibitem{propublica2}  Julia Angwin, Jeff Larson, Surya Mattu and Lauren Kirchner, ProPublica:  
  How We Analyzed the COMPAS Recidivism Algorithm, May 23, 2016, \url{https://www.propublica.org/article/how-we-analyzed-the-compas-recidivism-algorithm}. Cited 24 Oct 2018
%
\bibitem{CambridgeAnalytica} Nicholas Confessore,  New York Times, Cambridge Analytica and Facebook: The Scandal and the Fallout So Far, 4 Apr 2018, \url{https://www.nytimes.com/2018/04/04/us/politics/cambridge-analytica-scandal-fallout.html}. Cited 6 Nov 2018.
\bibitem{CambridgeAnalytica2} FAZ, Wir dachten, wir tun etwas völlig Normales, 21 Mar 2018, \url{http://www.faz.net/aktuell/wirtschaft/diginomics/skandal-um-cambridge-analytica-dachten-wir-tun-etwas-voellig-normales-15506137.html}
%
\bibitem{NetzDG} Bundesministerium der Justiz und für Verbraucherschutz,  Gesetz zur Verbesserung der Rechtsdurchsetzung in sozialen Netzwerken (Netzwerkdurchsetzungsgesetz - NetzDG), 1.9.2017, \url{https://www.gesetze-im-internet.de/netzdg/BJNR335210017.html}. Cited 6 Nov 2018.
%
\bibitem{facebook:restrictions} Mike Schroepfer, CTO Facebook, An Update on Our Plans to Restrict Data Access on Facebook, 4 Apr 2018,  \url{https://newsroom.fb.com/news/2018/04/restricting-data-access/}. Cited 6 Nov 2018.
%
\bibitem{laidoff} Ibrahin Diallo, The machine fired me, 17 June 2018, \url{https://idiallo.com/blog/when-a-machine-fired-me}. Cited 2 Nov 2018
%
\bibitem{reuters:amazon} Jeffrey Dastin, Amazon scraps secret AI recruiting tool that showed bias against women, Reuters News, 22 Oct 2018, 5:12 am, \url{https://www.reuters.com/article/us-amazon-com-jobs-automation-insight/amazon-scraps-secret-ai-recruiting-tool-that-showed-bias-against-women-idUSKCN1MK08G}. Cited 1 Nov 2018
%
\bibitem{MS:Tay}  Sarah Perez, Microsoft silences its new A.I. bot Tay, after Twitter users teach it racism, 2016,
\url{https://techcrunch.com/2016/03/24/microsoft-silences-its-new-a-i-bot-tay-after-twitter-users-teach-it-racism/}. Cited 1 Nov 2018
%
\bibitem{gmds2017}  Deutschen Gesellschaft für Medizinische Informatik, Biometrie und Epidemiologie (GMDS), Arne Manzeschke, Alfred Winter, Christoph Isele, Thomas Deserno, Frank Pallas, Karsten Weber and W. Niederlag , Workshop: Ethische Leitlinien wissenschaftlicher Fachgesellschaften, 4 May 2017, \url{https://gmds.de/ueber-uns/organisation/praesidiumskommissionen/ethische-fragen-in-der-medizinischen-informatik-biometrie-und-epidemiologie/}. Cited 6 Nov 2018.
\bibitem{forbes} Bernard Marr, Forbes, How Big Data Is Changing Insurance Forever,  16 Dec 2015, \url{https://www.forbes.com/sites/bernardmarr/2015/12/16/how-big-data-is-changing-the-insurance-industry-forever/}
%
\bibitem{James2018} Laura James, Oaths, pledges and manifestos: a master list of ethical tech values, doteveryone \url{https://doteveryone.org.uk}, 7 Mar 2018, \url{https://medium.com/doteveryone/oaths-pledges-and-manifestos-a-master-list-of-ethical-tech-values-26e2672e161c}. Cited 6  Nov 2018.
\bibitem{Erickson2018} Lucy C. Erickson, Natalie Evans Harris, and Meredith M. Lee, It's Time to Talk About Data Ethics, 26 Mar 2018, \url{https://www.datascience.com/blog/data-ethics-for-data-scientists}. Cited 6 Nov 2018.
\bibitem{sperrfrist:wahlen} Wissenschaftlicher Dienst des Bundestags Fachbereich WD 3: Verfassung und Verwaltung, Veröffentlichung der Ergebnisse von Umfragen vor Wahlen (Deutschland und Mitgliedstaaten der EU), Aktenzeichen WD 3 - 3000 - 058/18,  2018, \url{https://www.bundestag.de/blob/556748/ea25753e1c4a357a2c2c1c4791d4c4a8/wd-3-058-18-pdf-data.pdf}. Cited 2 Nov 2018.
%
\bibitem{oesterreich} Alexander Fanta, Österreichs Jobcenter richten künftig mit Hilfe von Software über Arbeitslose, 13 Oct 2018, \url{https://netzpolitik.org/2018/oesterreichs-jobcenter-richten-kuenftig-mit-hilfe-von-software-ueber-arbeitslose/}. Cited 2 Nov 2018.
%
\bibitem{panoptykon} Panoptykon Foundation, Jędrzej Niklas, Karolina Sztandar-Sztanderska and Katarzyna Szymielewicz, 2015, Warsaw, \url{https://panoptykon.org/sites/default/files/leadimage-biblioteka/panoptykon_profiling_report_final.pdf}. Cited 2 Nov 2018.
\bibitem{eCall} European Commission, Cybersecurity \& Digital Privacy Policy (Unit H.2), eCall: Time saved = lives saved, 14 Feb 2018, \url{https://ec.europa.eu/digital-single-market/en/ecall-time-saved-lives-saved}. Cited 7 Nov 2018.
\bibitem{zweig} Katharina Zweig, Wo Maschinen irren können, Impuls Algorithmenethik \#4, 2018, Bertelsmann Stiftung, doi: \url{https://doi.org/10.11586/2018006}
\bibitem{ASA:Guidelines} American Statistical Association,Ethical Guidelines for Statistical Practice, Approved April 2018, \url{https://www.amstat.org/ASA/Your-Career/Ethical-Guidelines-for-Statistical-Practice.aspx}. Cited 9 Nov 2018.
\bibitem{ACM:Guidelines} Association for Computing Machinery (ACM), ACM Code of Ethics and Professional Conduct, Approved June 2018, \url{https://www.acm.org/code-of-ethics}. Cited 9 Nov 2018
\bibitem{GI:Guidelines} Gesellschaft für Informatik, Ethical Guidelines of the German Informatics Society, 29 June 2018, \url{https://gi.de/ethicalguidelines/}. Cited 6 Nov 2018.
\bibitem{futurezone} Futurezone, AMS-Chef: "Mitarbeiter schätzen Jobchancen pessimistischer ein als der Algorithmus", 12 Oct 2018, \url{https://futurezone.at/netzpolitik/ams-chef-mitarbeiter-schaetzen-jobchancen-pessimistischer-ein-als-der-algorithmus/400143839}. Cited 6 Nov 2018.
%
\bibitem{ONeil} Cathy O'Neil, Weapons of Math Destruction: How Big Data Increases Inequality and Threatens Democracy,
 2016, Crown Publishing Group, New York, NY, USA 
%
\bibitem{TimoAiraksinen} Timo Airaksinen, The Philosophy of Professional Ethics, 2009,  in INSTITUTIONAL ISSUES INVOLVING ETHICS AND JUSTICE, edited Robert Charles Elliot, Vol 1, Page Number (201), in Encyclopedia of Life Support Systems (EOLSS), Developed under the Auspices of the UNESCO, Eolss Publishers, Paris, France, \url{http://www.eolss.net}
%
\bibitem{PollardJohnson} Pollard, T. J. and Johnson, A. E. W., The MIMIC-III Clinical Database \url{http://dx.doi.org/10.13026/C2XW26} (2016).
%
\bibitem{Pottegard2018} {Potteg{\r a}rd, Anton and Kristensen, Kasper Bruun and Ernst, Martin Thomsen and Johansen, Nanna Borup and Quartarolo, Pierre and Hallas, Jesper},
	{Use of N-nitrosodimethylamine (NDMA) contaminated valsartan products and risk of cancer: Danish nationwide cohort study},
	vol. 362 BMJ, 2018,	doi: 10.1136/bmj.k3851, BMJ Publishing Group Ltd, \url{https://www.bmj.com/content/362/bmj.k3851} 
%
\bibitem{Goldbergeretal} Goldberger AL, Amaral LAN, Glass L, Hausdorff JM, Ivanov PCh, Mark RG, Mietus JE, Moody GB, Peng C-K and Stanley HE. PhysioBank, PhysioToolkit, and PhysioNet: Components of a New Research Resource Complex Physiologic Signals. Circulation 101(23):e215-e220 [Circulation Electronic Pages; \url{http://circ.ahajournals.org/content/101/23/e215.full}]; 2000 (June 13).
%
\bibitem{HIPAA} U.S. Department of Health and Human Services. Standards for privacy of individually identifiable health information, final rule. Federal Register. 2002;45 CFR:160–164. [Ref list]
%
\bibitem{DAK1} Prof. Dr. Wolfgang Greiner, Manuel Batram, Oliver Damm, Stefan Scholz, and Julian Witte, 2018, Kinder- und Jugendreport 2018, Beiträge zur Gesundheitsökonomie und Versorgungsforschung (Band 23), Andreas Storm (Herausgeber), DAK-Gesundheit
%
\bibitem{DAK2} Dr. Benjamin Kuntz, Elvira Mauz und PD Dr. Thomas Lampert, Die KiGGS-Studie des Robert Koch-Instituts: Studiendesign, Erhebungsinhalte und Ergebnisse zur gesundheitlichen Ungleichheit im Kindes- und Jugendalter  – Robert Koch-Institut, Berlin, Kinder- und Jugendreport 2018, Beiträge zur Gesundheitsökonomie und Versorgungsforschung (Band 23), Andreas Storm (Herausgeber), DAK-Gesundheit
%
\bibitem{KIGGS} Bärbel-Maria Kurth, Panagiotis Kamtsiuris, Heike Hölling, Martin Schlaud, Rüdiger Dölle, Ute Ellert, Heidrun Kahl, Hiltraud Knopf, Michael Lange, Gert BM Mensink, Hannelore Neuhauser, Angelika Schaffrath Rosario, Christa Scheidt-Nave, Liane Schenk, Robert Schlack, Heribert Stolzenberg, Michael Thamm, Wulf Thierfelder and Ute Wolf, The challenge of comprehensively mapping children's health in a nation-wide health survey: Design of the German KiGGS-Study BMC Public Health20088:196 \url{https://doi.org/10.1186/1471-2458-8-196}  Kurth et al; licensee BioMed Central Ltd. 2008
%
\bibitem{Pariser:FilterBubble} Eli, Pariser,
 The filter bubble: how the new personalized web is changing what we read and how we think,
 Penguin Books, 2012, New York, N.Y., ISBN 0143121235
%
\bibitem{Horner} Jennifer Horner, Ph.D., J.D, 2003. Morality, Ethics, and Law: Introductory Concepts. SEMINARS IN SPEECH AND LANGUAGE. Vol 24 (4) 263-274.
\bibitem{CNIL2017}  Commission Nationale Informatique \& Liberte, HOW CAN HUMANS KEEP THE UPPER HAND? The ethical matters raised by algorithms and artificial intelligence, Dec 2017, \url{https://www.cnil.fr/en/how-can-humans-keep-upper-hand-report-ethical-matters-raised-algorithms-and-artificial-intelligence}. Cited 2 Nov 2018.
\bibitem{CNIL2018}  Commission Nationale Informatique \& Liberte, Algorithms and artificial intelligence: CNIL’s report on the ethical issues, 25 May 2018, \url{https://www.cnil.fr/en/algorithms-and-artificial-intelligence-cnils-report-ethical-issues}. Cited 2 Nov 2018. 
\end{thebibliography}
\end{document}